\documentclass[prb,showpacs,showkeys,showpacs,twocolumn,unsortedaddress]{revtex4}
\usepackage{amsmath,amssymb,epic,eepic}
\usepackage{siunitx}
\usepackage[utf8]{inputenc}
\usepackage{color}
\usepackage{epsfig}
\usepackage{bm}
\usepackage{tikz}
\usepackage{datetime}
\usetikzlibrary{shadings,fadings,circuits,circuits.ee,
	circuits.ee.IEC}

\IfFileExists{srcltx.sty}{\usepackage[active]{srcltx}}

\newcommand{\me}{\mathrm{e}}
\newcommand{\mi}{\mathrm{i}}
\newcommand{\md}{\mathrm{d}}

\DeclareMathOperator{\trace}{tr}

\begin{document}

\title{
Perturbative fluctuation dissipation relation for non-equilibrium finite frequency noise in quantum circuits
}

\author{Benjamin Roussel$^{1}$, Pascal Degiovanni$^{1}$ and In\`es Safi$^{2}$}

\email{safi@lps.u-psud.fr}

\affiliation{(1) Universit\'e de Lyon, F\'ed\'eration de Physique A.-M. Amp\`ere,\\
CNRS - Laboratoire de Physique de l'Ecole Normale Sup\'erieure de Lyon,\\
46 All\'ee d'Italie, 69364 Lyon Cedex 07, France}

\affiliation{
(2) Laboratoire de Physique des Solides (UMR 5802),\\ Universit\'e
Paris-Sud,
Batiment 510, 91405 Orsay, France
}

\begin{abstract} 
We develop a general perturbative computation of finite-frequency quantum noise
which applies, in particular, to both good or weakly transmitting strongly
correlated conductors coupled to a generic environment.
Under a minimal set of hypotheses, we show that the noise
can be expressed through the non-equilibrium DC current
only, generalizing a non-equilibrium fluctuation
dissipation relation. We use this relation to derive
explicit predictions for the non equilibrium finite frequency
noise for a single channel conductor connected to an arbitrary Ohmic
environment. 
\end{abstract}



\maketitle
 
In a quantum circuit composed of several coherent conductors,
electronic transport 
depend on the global circuit even when conductors are
separated by 
distances greater than
the electronic coherence length as, for instance,
in the dynamical Coulomb blockade (DCB)~\cite{Devoret:1990-1,Girvin:1990-1}. Consequently,
classical laws of electricity such as 
the impedance composition law are violated~\cite{Altimiras:2007-1}.
It is therefore important to look for
quantum laws of electricity replacing the classical ones. They must be
independent on details of the dynamics 
such as coupling to the environment or 
screened Coulomb interactions within
each conductor. 
For example, the recently derived 
universal 
relation between current correlations and generalized admittances 
for non-linear
time dependent transport in quantum circuits~\cite{Safi:2011-1}
is a consistancy condition valid independently of
the details of the system's initial density matrix, Hamiltonian and coupling to its environment.

Assuming thermal equilibrium also leads to system independent relations such as
the standard fluctuation dissipation
theorem (FDT)~\cite{Callen:1951-1,Kubo:1957-1} derived
within a linear response theory.
This FDT is now embraced by a general corpus of non-equilibrium 
fluctuation relations derived 
for charge transport in the classical
regime~\cite{Altland:2010-1,Altland:2010-2},
non-linear DC transport through a single quantum 
conductor~\cite{Forster:2008-1} and quantum circuits in the limit
of weak environmental effects~\cite{Tobiska:2005-1}.
However the validity of such results for quantum circuits involving strong 
environmental effects is still an open question. 

Such effects have been studied within the Dynamical Coulomb Blockade
problem, first in a tunnel junction 
coupled to a linear environment~\cite{Ingold:1992-1}.
The case of a good conductor coupled to a small impedance
was then considered~\cite{Golubev:2001-1,Levy-Yeyati:2001-1,Kindermann:2003-1} and
further works~\cite{Safi:2004-1,Jezouin:2013-1,Golubev:2005-1} have
completed our understanding of the DCB of the current for higher impedances. 
The question of the DCB of the noise has been explored
theoretically~\cite{Kindermann:2004-1,Safi:2004-1,Zamoum:2012-1},
leading to experimental investigations~\cite{Altimiras:2014-1}.
The importance of fluctuation dissipation relations (FDRs) 
relating the finite frequency (FF) quantum noise to the DC
non-equilibrium average current progressively emerged~\cite{Safi:2014-1,Parlavecchio:2015-1}.
In fact, similar FDRs 
had been
derived in the stationary regime in 
the fractional quantum Hall effect~\cite{Bena:2007-1} and, for 
the symmetrized FF noise for free
quasiparticles~\cite{Dahm:1969-1,Rogovin:1974-1},
in presence of 
a linear environment~\cite{Lee:1996-2} and
in presence of arbitrary
interactions~\cite{Sukhorukov:2001-1}. 

In this Letter, we show that the
FDR between the FF quantum noise and DC non-equilibrium current~\cite{Safi:2014-1} is 
valid independently of the
details of the conductor as well as of its environment provided the
following hypotheses are satisfied~:
\textit{(i)} validity and finiteness of perturbation theory, \textit{(ii)} absence of superconducting 
current and \textit{(iii)} detailed balance in the limit of 
vanishing tunneling, a condition to be obeyed by the total tunneling operator, which includes phase
fluctuations.

This FDR can be exploited in various ways: first,
it provides a test of the hypotheses \textit{(i)}-\textit{(iii)} when 
one 
measures independantly both the average current and its FF noise. 
Secondly, it provides explicit predictions for the FF
noise from the DC non-equilibrium current which is easier to compute
than the noise.
As an illustration, using the mapping between the DCB problem and the 
Tomonaga-Luttinger liquid (TLL) theory~\cite{Safi:2001-1,Jezouin:2013-1}, our FDR
gives explicit predictions for the effect of an Ohmic environment on the FF noise from that on
the DC non-equilibrium current, thus extending 
previous~\cite{Altimiras:2014-1,Bena:2007-1,Zamoum:2012-1,Souquet:2013-1} works.

\medskip

\begin{figure}[htbp] 
	\centering
\begin{tikzpicture}[circuit ee IEC]
	\def\condwidth{0.7}
	\begin{scope}[shift={(1,1.5)}]
		\coordinate (entercond) at (0,0);
		\coordinate (exitcond) at (\condwidth+0.04+\condwidth,0);
		\draw[semithick] (0,-0.3) rectangle (\condwidth,0.3);
		\draw[shift={(\condwidth+0.04,0)},semithick] (0,-0.3) rectangle (\condwidth,0.3);
		\draw (0.7*\condwidth,0) circle (0.04);
		\node[left] at (0.7*\condwidth,0) {$q$};
		\draw[-stealth] (0.7*\condwidth+0.04,0) -- (0.3*\condwidth+\condwidth+0.04,0);
	\end{scope}
	\node[draw,rectangle] (env) at (1+\condwidth+0.02,0) {Environment};
	\draw (entercond) -- (0,1.5)
		to[battery={info'={$V$}}]
		(0,0) to (env);
	\draw (exitcond) -- ++(1,0) -- ++(0,-1.5) to (env);
\end{tikzpicture}
	\caption{
		A general two terminal conductor 
		is embedded into a quantum circuit in which
		it is coupled to an impedance $Z(\omega)$ and/or other conductors 
		constituting its environment. Here the conductor is a spatially extended tunnel
		junction with capacitive couplings.
	}\label{setup}
\end{figure}
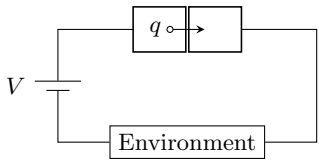

We consider a quantum circuit built from 
a two terminal conductor with possible strong correlations
and coupled to an arbitrary environment involving other conductors 
without any restriction
(see Figure
\ref{setup}). 

The circuit is biased by a DC voltage $V$ and its Hamiltonien is decomposed into
\begin{align}
	\label{Hamiltonian} 
	\mathcal{H}(t) &= \mathcal{H}_0 
	+\me^{\mi\omega_{\mathrm{dc}}
	t}\mathcal{A}+\me^{-\mi\omega_{\mathrm{dc}} t}\mathcal{A}^{\dagger}\,.
\end{align} 
Here 
$\mathcal{H}_0$ involves internal or mutual Coulomb interactions as well as the
environment. 
Charge transfer processes across the junction are encoded within tunneling operators $\mathcal{A}$ 
and $\mathcal{A}^\dagger$ which 
include environment induced quantum fluctuations of the phase jump across the conductor.
The DC bias $V$ introduces a frequency scale $\omega_{\mathrm{dc}}=qV/\hbar$ where
$q$ denotes the effective renormalized charge of transferred quasi-particles.
The current operator is defined as:
\begin{equation}
	\label{eq:current}
	\hat{I}(t) = 
	\frac{\mi q}{\hbar}\left(
		\me^{\mi\omega_{\mathrm{dc}} t} \mathcal{A}
		- \me^{-\mi\omega_{\mathrm{dc}} t} \mathcal{A}^{\dagger}
	\right)\,.
\end{equation} 
Here we focus on FF noise expressed perturbatively with respect
to $\mathcal{A}$. Several approaches to this problem are possible. One could 
model the electromagnetic environment
in a quantum input/output
approach, a first step in this direction being taken by Parlavecchio 
\textit{et al} for a tunnel junction coupled to an LC oscillator~\cite{Parlavecchio:2015-1}. 
On the other hand, the standard field 
theoretical approach followed here 
considers an adiabatic branching of tunneling from an initial condition at 
$t\rightarrow -\infty$ described by a density operator $\hat{\rho}_0$.
Since
$\hat{I}(t)$ is of first order with respect to $\mathcal{A}$, averages can be
computed with respect to $\hat{\rho}_0$ and using the Heisenberg
representation with respect to $\mathcal{H}_0$:
\begin{equation} 
	\mathcal{A}_0(t) =
	\me^{\mi\mathcal{H}_0 t} \mathcal{A}\,\me^{-\mi\mathcal{H}_0t}
	\label{eq:heisenberg} 
\end{equation} 
To ensure absence of a super current
in the limit of vanishing tunneling whenever superconductors are present,
we require that:
\begin{equation}
	\label{condition} 
	\langle \mathcal{A}_0(t)\mathcal{A}_0(0) \rangle_0=0,
\end{equation} 
where $\langle \ldots\rangle_0 = \trace\left[\hat{\rho}_0 ...\right]$. 
At the lowest order, the FF noise is given by
\begin{equation}
	\label{Sdefinition}
	S(\omega,\omega_{\mathrm{dc}}) = 
	\int \me^{\mi\omega (t-t')}
	\langle \hat I_0(t')\hat I_0(t)\rangle_0\text{d}t'
\end{equation} 
where $\hat{I}_0(t)$ is obtained by replacing $\mathcal{A}$ in Eq.~\eqref{eq:current}
by $\mathcal{A}_0(t)$ given by Eq.~\eqref{eq:heisenberg}.
The detailed balance (hypothesis \textit{(iii)})
constrains the occupation probabilities of the many body eigenstates of the circuit 
in the limit of vanishing tunneling to be given by
a thermal distribution $\rho_0$ with effective temperature $T$.
This 
enables us to relate 
the FF quantum noise to
the DC non-equilibrium current across the circuit (see Appendix~\ref{appendix:FDR}):
\begin{align}
		\label{eq:fdr:dc}
		S(\omega,\omega_{\mathrm{dc}})= q \left[
		N(\omega_{\mathrm{dc}}+\omega)I(\omega_{\mathrm{dc}}+\omega)\right.\nonumber\\
		\left. +(1+
		N(\omega_{\mathrm{dc}}-\omega))I(\omega_{\mathrm{dc}}-\omega)\right]\,
\end{align} 
where $I(\omega_{\text{dc}})$ denotes the non equilibrium dc current at bias voltage $V$ and 
$N(\omega)$ denotes the Bose occupation number 
at temperature $T$.  
This perturbative non-equilibrium FDR is the central result of
this Letter. It is model independant and its validity solely relies
on hypotheses \textit{(i)}-\textit{(iii)}.
Note that Eq.~\eqref{eq:fdr:dc} also implies the 
FDR 
for the symmetrized noise previously derived 
in more specific contexts~\cite{Rogovin:1974-1,Lee:1996-2,Sukhorukov:2001-1}.

\medskip

Let us now analyze the various regimes of the FF noise 
$S(\omega,\omega_{\mathrm{dc}})$ as a function of
$\omega$ and $\omega_{\mathrm{dc}}$ at fixed temperature $T$.
First of all, its asymmetry with respect to $\omega$ 
is related to the
non-equilibrium admittance
$G(\omega,\omega_{\mathrm{dc}})$~\cite{Safi:2011-1}. As a consequence, both
quantities
have a perturbative expression in terms of the 
non-equilibrium DC current~\cite{Safi:2010-1}:
\begin{subequations}
	\label{FDRA}
\begin{align} 
	S(-\omega,\omega_{\mathrm{dc}}) - S(\omega,\omega_{\mathrm{dc}}) 
	 = 2\hbar\omega \,\Re\left(G(\omega,\omega_{\mathrm{dc}})\right)
	\label{admittance}\\ 
	 = 
	q\left\{
		I(\omega_{\mathrm{dc}}+\omega)-I(\omega_{\mathrm{dc}}-\omega)
	\right\}\,. 
\end{align}
\end{subequations}
Secondly, the noise $S(\omega,\omega_{\mathrm{dc}})$ is
even with respect to $\omega_{\mathrm{dc}}$ whenever
particle-hole symmetry holds: in that case, 
a spectral decomposition shows that the DC characteristic 
is odd. 
Generically
$I(\omega_{\mathrm{dc}})$ has the same sign as
$\omega_{\mathrm{dc}}$ and $I(\omega_{\mathrm{dc}}=0)=0$. 

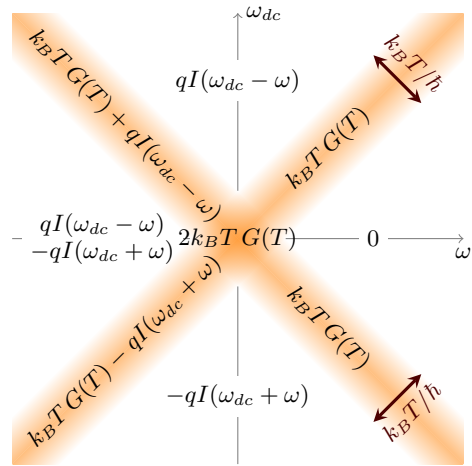
\begin{figure}[htbp] 
	\begin{center}
\begin{tikzpicture}
	\def\maxwidth{6}
	\pgfmathsetmacro{\xmax}{0.5*\maxwidth}
	\pgfmathsetmacro{\ymax}{\xmax}
	\pgfmathsetmacro{\Twide}{\maxwidth*0.075}
	\tikzfading[name=fadea, left color=transparent!50, right color=transparent!100]
	\tikzfading[name=fadeb, right color=transparent!50, left color=transparent!100]
	\tikzfading[name=fadec, top color=transparent!100, bottom color=transparent!50]
	\tikzfading[name=faded, bottom color=transparent!100, top color=transparent!50]

	\draw[gray,->] (-\xmax, 0) -- (-\Twide*1.4142, 0)
		(\Twide*1.4142, 0) -- (\xmax,0)
		node[below,black] {$\omega$};
	\draw[gray,->] (0,-\ymax) -- (0, -\Twide*1.4142)
		(0, \Twide*1.4142) -- (0, \ymax)
		node[right,black] {$\omega_{dc}$};
	\node[fill=white,align=center] (l) at (0.6*\xmax,0) {$0$};
	\node[fill=white,align=center] (r) at (-0.6*\xmax,0)
		{$qI(\omega_{dc}-\omega)$\\ $-qI(\omega_{dc}+\omega)$};
	\node[fill=white,align=center] (a) at (0,0.7*\ymax)
		{$qI(\omega_{dc}-\omega)$};
	\node[fill=white,align=center] (b) at (0,-0.7*\ymax)
		{$-qI(\omega_{dc}+\omega)$};

	\begin{scope}
		\clip (-\xmax, -\ymax) rectangle (\xmax, \ymax);
		\begin{scope}[transform canvas={rotate=45}]
			\fill[path fading=fadec,orange]
				(-1.4142*\ymax, 0) rectangle (1.4142*\xmax, \Twide);
			\fill[path fading=faded,orange]
				(-1.4142*\ymax, -\Twide) rectangle (1.4142*\xmax, 0);
			\fill[path fading=fadea,orange]
				(0, -1.4142*\ymax) rectangle (\Twide, 1.4142*\xmax);
			\fill[path fading=fadeb,orange]
				(-\Twide, -1.4142*\ymax) rectangle (0, 1.4142*\xmax);
			\node (ar) at (0.4*1.4142*\xmax,0) {$k_BT\,G(T)$};
			\node[rotate=-90] (br) at (0,-0.4*1.4142*\xmax) {$k_BT\,G(T)$};
			\node[rotate=-90] (al) at (0,0.5*1.4142*\xmax)
				{$k_BT\,G(T)+qI(\omega_{dc}-\omega)$}; 
		\end{scope}
		\node[rotate=45] (bl) at (45:-0.5*1.4142*\xmax)
			{$k_BT\,G(T)-qI(\omega_{dc}+\omega)$}; 
		\node (m) at (0,0) {$2k_BT\,G(T)$};
	\end{scope}

	\begin{scope}[rotate=45,shift={(0:1*\xmax)}]
		\draw[stealth-stealth, line width=1.2pt,
			red!30!black]
			(0,-\Twide) -- (0,\Twide)
			node[midway, above, rotate=-45] {$k_B T / \hbar$};
	\end{scope}
	\begin{scope}[rotate=-45,shift={(0:1*\xmax)}]
		\draw[stealth-stealth, line width=1.2pt,
			red!30!black]
			(0,-\Twide) -- (0,\Twide)
			node[midway, below, rotate=45] {$k_B T / \hbar$};
	\end{scope}
\end{tikzpicture}
		\caption{
\label{fig:quadrants:noise} (Color online) Various regimes for the
			noise as a function of $\omega$ and $\omega_{\mathrm{dc}}$. The
			$(\omega,\omega_{\mathrm{dc}})$ plane is partitioned into four quadrants
			separated by diagonal bands of width $k_BT/\hbar$ (orange)
			corresponding to the typical energy scale of thermal electron/hole
			pairs.  The quantum regime corresponds to $\hbar|\omega|\gg k_BT$
			or $\hbar|\omega_{\mathrm{dc}}|\gg k_BT$ whereas the thermal regime corresponds to
			both $\hbar\omega$ and $\hbar\omega_{\mathrm{dc}}$ much smaller than $k_BT$.
		}
	\end{center} 
\end{figure}

Being mostly interested in the quantum regime, we shall explore frequencies
$\omega_{\mathrm{dc}}$ and $\omega$ well above the thermal scale
$k_BT/\hbar$. Since $\hbar\omega$ represents the energy of photons emitted
($\omega>0$) or absorbed ($\omega<0$) by the conductor, it is natural to
compare it to the energy scale $|qV|$.
This leads us to partition the $(\omega,\omega_{\mathrm{dc}})$ plane into four
quadrants (see Fig.~\ref{fig:quadrants:noise})
separated by diagonal bands $\hbar|\omega\pm\omega_{\mathrm{dc}}|\lesssim k_BT$
in which thermal fluctuations turn out to play a role even in the quantum regime.

We first look at the physics far from these 
bands.
In the $\omega >|\omega_{\mathrm{dc}}|$ quadrant (see 
Fig.~\ref{fig:quadrants:noise}), the system cannot emit any photon
(at first order in perturbation theory). Therefore, we expect the FF
noise to vanish~\cite{Safi:2011-1}:
$S(\omega,\omega_{\mathrm{dc}})=0$. Then,
Eq.~\eqref{FDRA} leads to the expression of the FF
noise in the $\omega<-|\omega_{\mathrm{dc}}|$ quadrant where it represents the 
ability of the circuit to absorb radiation. It is then naturally related
to the dissipative part of the non equilibrium admittance (see Eq.~\eqref{FDRA}): 
$S(\omega,\omega_{\mathrm{dc}})=2\hbar\omega
\Re{\left(G(\omega,\omega_{\mathrm{dc}})\right)}$. 
When particle-hole symmetry holds, 
$S(-\omega,0)=2qI(\omega)$ which reduces to the
usual expression for the FF equilibrium noise only for a linear system.

We now consider
the off-diagonal
quadrants $|\omega_{\mathrm{dc}}|>|\omega|$:
the vanishing of Bose occupation numbers there implies that
the FF noise is proportional $I(\omega_{\text{dc}}-\omega)$
(see Fig.~\ref{fig:quadrants:noise}).
In particular, at zero frequency and for
$|qV|\gg k_BT$, the noise has a Poissonian expression:
\begin{equation}
	\label{DC_noise_DC}
	S(0,\omega_{\mathrm{dc}})=qI(\omega_{\mathrm{dc}})\,
\end{equation} 
shown here to be
valid for an extended and interacting tunneling region~\cite{Safi:1999-2,Chevallier:2010-1}
with a generic environment whereas it was
originally derived for an isolated tunnel junction
between decoupled conductors~\cite{Levitov:2004-1}.

Let us then look into diagonal bands where 
thermal fluctuations
generate neutral excitations with energies below $k_BT$.
In the thermal regime, when both $\omega$ and $\omega_{\mathrm{dc}}$ are smaller than
$k_BT/\hbar$, we recover the Johnson Nyquist~\cite{Nyquist:1928-1}
noise $S(0,0)=2k_BTG(T)$ where
$G(T)=G(\omega=0,\omega_{\mathrm{dc}}=0,T)$ is the linear conductance which
may depend on temperature.
Remarquably, at positive frequencies, in the quantum regime, the FF
noise along the diagonals is half of the equilibrium noise whereas for negative frequencies, 
it picks up an extra non-equilibrium contribution $\pm qI(\omega_{\mathrm{dc}}-\omega)$.

At finite frequency,
one usually measures
the differential of the FF noise 
with respect to the DC voltage or 
equivalently $\omega_{\mathrm{dc}}$. In particular, 
at low frequency,
Eq.~\eqref{eq:fdr:dc} implies that
this differential noise is related to the differential conductance
for $|qV|\ll k_BT$: 
\begin{equation}
	 \label{zerof_differential_noise_DC_non}
\left(\frac{\partial S}{\partial\omega_{\mathrm{dc}}}
\right)_{\hbox to 0pt{$\scriptstyle \omega=\omega_{\mathrm{dc}}=0$}}(T)
	= 
	k_BT 
\left(\frac{\partial G}{\partial\omega_{\mathrm{dc}}}
\right)_{\hbox to 0pt{$\scriptstyle \omega=\omega_{\mathrm{dc}}=0$}}(T).
\end{equation}
This relation has been derived for weak environmental effects~\cite{Tobiska:2005-1} 
and for an isolated interacting quantum conductor 
within a Hartree framework~\cite{Forster:2008-1}.
It has been
experimentally tested in quantum dots~\cite{Nakamura:2010-1}.

To illustrate our general result, we will now use
Eq.~\eqref{eq:fdr:dc} to 
obtain explicit predictions for the FF noise of a
single channel conductor
in series with an Ohmic impedance $R$.
As shown theoretically~\cite{Safi:2004-1} and confirmed 
experimentally~\cite{Jezouin:2013-1}, this situation 
can be described in terms of 
a localized barrier in a Luttinger liquid (TLL)~\cite{Kane:1992-1,Saleur:1998-1} with
interaction parameter 
$K=(1+R/R_q)^{-1}<1$ where $R_q=h/e^2$. A stronger coupling to the environment thus
corresponds to stronger repulsive interactions
in the TLL.
The tunneling regime ($R_qG\ll 1$) of the DCB problem corresponds to a
strong barrier in the TLL, \textit{i.e} to the vicinity of its IR fixed point 
($K<1$) whereas
a good conductor ($R_qG\simeq 1$) corresponds to 
the case of a weak barrier, \textit{i.e.} to the vicinity of its unstable UV 
fixed point~\cite{Saleur:1998-1}.

Consequently, for a good conductor, we expect perturbative
results to be valid when the largest energy among $|qV|$ and
$k_BT$ is greater than $E_B$, an
intrinsic energy of this impurity problem scaling
as $E_B\simeq \hbar\omega_{\mathrm{RC}}(1-\mathcal{T})^{1/2(1-K)}$ in terms
of 
$\mathcal{T}= R_qG(R/R_q=0)\lesssim 1$, the bare transmission (no DCB) of 
the conductor~\cite{Safi:2004-1,Jezouin:2013-1} and of
the UV cutoff of the problem $\omega_{\mathrm{RC}}$.
A linear DC characteristic 
$V=(R+R_q)I$ corresponding to the series addition of 
resistances is recovered when $k_BT\gg
|qV|$ and $\mathrm{max}(k_BT,|qV|)\gg E_B$.
For a good conductor, voltage division within the circuit then leads to
a charge renormalization $q=-eR_q/(R+R_q)=-eK$.
When $|qV|\gg k_BT$, keeping $|qV|\gg E_B$,
the DC characteristic is no longer linear: $I(V,T)=KV/R_q-I_B(V,T)$ where
$I_B(V,T)$ is the weak backscattering current.
Predictions
for the backscattering noise $S_B(\omega,\omega_{\mathrm{dc}})$
follow from Eq.~\eqref{eq:fdr:dc} and from the perturbative expression~\cite{Chamon:1995-1}:
\begin{equation}
\label{eq:Chamon}
I_B(V,T) =  V\,G_\Delta(T)\,
\frac{\sinh{\left(\frac{qV}{2k_BT}\right)}}{\frac{qV}{2k_BT}}
\frac{\left|\Gamma\left(\Delta+\frac{\mi qV}{2\pi k_BT}\right)\right|^2}{\Gamma(\Delta)^2} 
\end{equation}
where
$\Delta$ is 
equal to $K<1$, $q=-Ke$ and $G_\Delta(T)$ is a backscattering
conductance scaling as $G_\Delta(T)/G_{\Delta=1}(T)=
(\Gamma(\Delta)^2/\Gamma(2\Delta))\times(\hbar\omega_{\mathrm{RC}}/\pi
k_BT)^{2(1-\Delta)}$ (see Appendix \ref{appendix:TLL}).
In fact, Eq.~\eqref{eq:Chamon} is valid as long as $\text{max}(|qV|,k_BT)>E_B$.
Consequently, for $k_BT\gg E_B$, Eq.~\eqref{eq:Chamon} can be used in Eq.~\eqref{eq:fdr:dc}
without further restrictions on $(\omega,\omega_{\mathrm{dc}})$ whereas, 
at lower temperatures
($k_BT\lesssim E_B$),  
our result for $S_B(\omega,\omega_{\mathrm{dc}})$ is valid
for
$\hbar|\omega_{\mathrm{dc}}\pm \omega|$ greater than $E_B$. 

\medskip

Remarkably, the mapping of the DCB on the TLL model also gives access to
the low energy behavior ($|qV|\ll E_B$ and $k_BT\ll
E_B$) of a good conductor. This behavior also describes the low energy physics of
a tunnel junction due to the strong DCB~\cite{Safi:2004-1}.
Perturbation theory can then be applied close to the IR fixed point corresponding to a disconnected
TLL. Since 
the total
effective resistance of the circuit is now much larger than $R$, 
no voltage division takes place thus giving $q=-e$.
Eq.~\eqref{eq:Chamon} can then be used to compute $I(V,T)$ with
$\Delta=K^{-1}=1+R/R_q>1$ and $G_\Delta(T)$ then corresponds to the linear conductance
$G(T)$.
Predictions for $S(\omega,\omega_{dc})$ from Eq.~\eqref{eq:fdr:dc} 
are now valid when $|\hbar\omega|\ll E_B$ and $|qV|\ll E_B$.
They apply as well to a weakly transmitting conductor where perturbation theory is expected
to be valid for all energy scales smaller than the cutoff energy 
$\hbar\omega_{\mathrm{RC}}$.

\medskip

To understand the effect of the environment on the FF noise in both regimes, we plot the 
ratio $\mathcal{S}_R(\omega,\omega_{\mathrm{dc}})$
of the noise
for $R\neq 0$ to its value at $R=0$ (no DCB).
We present our results assuming $\hbar\omega_{\mathrm{RC}}=40\,E_B$ and for two values of
$R$: $R=R_q/2$ ($K=2/3$) and 
$R=2R_q$ ($K=1/3$). Note that the FF noise is even with respect to
$\omega_{\mathrm{dc}}$ due to the electron/hole symmetry of the TLL model.

\begin{figure}[htbp] 
	\begin{center}
\begingroup
  \makeatletter
  \providecommand\color[2][]{%
    \GenericError{(gnuplot) \space\space\space\@spaces}{%
      Package color not loaded in conjunction with
      terminal option `colourtext'%
    }{See the gnuplot documentation for explanation.%
    }{Either use 'blacktext' in gnuplot or load the package
      color.sty in LaTeX.}%
    \renewcommand\color[2][]{}%
  }%
  \providecommand\includegraphics[2][]{%
    \GenericError{(gnuplot) \space\space\space\@spaces}{%
      Package graphicx or graphics not loaded%
    }{See the gnuplot documentation for explanation.%
    }{The gnuplot epslatex terminal needs graphicx.sty or graphics.sty.}%
    \renewcommand\includegraphics[2][]{}%
  }%
  \providecommand\rotatebox[2]{#2}%
  \@ifundefined{ifGPcolor}{%
    \newif\ifGPcolor
    \GPcolortrue
  }{}%
  \@ifundefined{ifGPblacktext}{%
    \newif\ifGPblacktext
    \GPblacktexttrue
  }{}%
  \let\gplgaddtomacro\g@addto@macro
  \gdef\gplbacktext{}%
  \gdef\gplfronttext{}%
  \makeatother
  \ifGPblacktext
    \def\colorrgb#1{}%
    \def\colorgray#1{}%
  \else
    \ifGPcolor
      \def\colorrgb#1{\color[rgb]{#1}}%
      \def\colorgray#1{\color[gray]{#1}}%
      \expandafter\def\csname LTw\endcsname{\color{white}}%
      \expandafter\def\csname LTb\endcsname{\color{black}}%
      \expandafter\def\csname LTa\endcsname{\color{black}}%
      \expandafter\def\csname LT0\endcsname{\color[rgb]{1,0,0}}%
      \expandafter\def\csname LT1\endcsname{\color[rgb]{0,1,0}}%
      \expandafter\def\csname LT2\endcsname{\color[rgb]{0,0,1}}%
      \expandafter\def\csname LT3\endcsname{\color[rgb]{1,0,1}}%
      \expandafter\def\csname LT4\endcsname{\color[rgb]{0,1,1}}%
      \expandafter\def\csname LT5\endcsname{\color[rgb]{1,1,0}}%
      \expandafter\def\csname LT6\endcsname{\color[rgb]{0,0,0}}%
      \expandafter\def\csname LT7\endcsname{\color[rgb]{1,0.3,0}}%
      \expandafter\def\csname LT8\endcsname{\color[rgb]{0.5,0.5,0.5}}%
    \else
      \def\colorrgb#1{\color{black}}%
      \def\colorgray#1{\color[gray]{#1}}%
      \expandafter\def\csname LTw\endcsname{\color{white}}%
      \expandafter\def\csname LTb\endcsname{\color{black}}%
      \expandafter\def\csname LTa\endcsname{\color{black}}%
      \expandafter\def\csname LT0\endcsname{\color{black}}%
      \expandafter\def\csname LT1\endcsname{\color{black}}%
      \expandafter\def\csname LT2\endcsname{\color{black}}%
      \expandafter\def\csname LT3\endcsname{\color{black}}%
      \expandafter\def\csname LT4\endcsname{\color{black}}%
      \expandafter\def\csname LT5\endcsname{\color{black}}%
      \expandafter\def\csname LT6\endcsname{\color{black}}%
      \expandafter\def\csname LT7\endcsname{\color{black}}%
      \expandafter\def\csname LT8\endcsname{\color{black}}%
    \fi
  \fi
  \setlength{\unitlength}{0.0500bp}%
  \begin{picture}(4760.00,2820.00)%
    \gplgaddtomacro\gplbacktext{%
      \csname LTb\endcsname%
      \put(1666,2587){\makebox(0,0){\strut{}$R = 2R_q$}}%
    }%
    \gplgaddtomacro\gplfronttext{%
      \csname LTb\endcsname%
      \put(953,463){\makebox(0,0){\strut{}-30}}%
      \csname LTb\endcsname%
      \put(1310,463){\makebox(0,0){\strut{}-15}}%
      \csname LTb\endcsname%
      \put(1666,463){\makebox(0,0){\strut{} 0}}%
      \csname LTb\endcsname%
      \put(2022,463){\makebox(0,0){\strut{} 15}}%
      \csname LTb\endcsname%
      \put(2379,463){\makebox(0,0){\strut{} 30}}%
      \csname LTb\endcsname%
      \put(1666,184){\makebox(0,0){\strut{}$\hbar \omega / E_B$}}%
      \csname LTb\endcsname%
      \put(582,705){\makebox(0,0)[r]{\strut{}-40}}%
      \csname LTb\endcsname%
      \put(582,1022){\makebox(0,0)[r]{\strut{}-20}}%
      \csname LTb\endcsname%
      \put(582,1339){\makebox(0,0)[r]{\strut{} 0}}%
      \csname LTb\endcsname%
      \put(582,1656){\makebox(0,0)[r]{\strut{} 20}}%
      \csname LTb\endcsname%
      \put(582,1973){\makebox(0,0)[r]{\strut{} 40}}%
      \csname LTb\endcsname%
      \put(175,1339){\rotatebox{-270}{\makebox(0,0){\strut{}$\hbar \omega_{\mathrm{dc}} / E_B$}}}%
      \csname LTb\endcsname%
      \put(952,2244){\makebox(0,0){\strut{} 0}}%
      \csname LTb\endcsname%
      \put(1309,2244){\makebox(0,0){\strut{} 5}}%
      \csname LTb\endcsname%
      \put(1666,2244){\makebox(0,0){\strut{} 10}}%
      \csname LTb\endcsname%
      \put(2023,2244){\makebox(0,0){\strut{} 15}}%
      \csname LTb\endcsname%
      \put(2380,2244){\makebox(0,0){\strut{} 20}}%
    }%
    \gplgaddtomacro\gplbacktext{%
      \csname LTb\endcsname%
      \put(3618,2587){\makebox(0,0){\strut{}$R=R_q/2$}}%
    }%
    \gplgaddtomacro\gplfronttext{%
      \csname LTb\endcsname%
      \put(2905,463){\makebox(0,0){\strut{}-30}}%
      \csname LTb\endcsname%
      \put(3262,463){\makebox(0,0){\strut{}-15}}%
      \csname LTb\endcsname%
      \put(3618,463){\makebox(0,0){\strut{} 0}}%
      \csname LTb\endcsname%
      \put(3974,463){\makebox(0,0){\strut{} 15}}%
      \csname LTb\endcsname%
      \put(4331,463){\makebox(0,0){\strut{} 30}}%
      \csname LTb\endcsname%
      \put(3618,184){\makebox(0,0){\strut{}$\hbar \omega / E_B$}}%
      \csname LTb\endcsname%
      \put(2534,705){\makebox(0,0)[r]{\strut{}}}%
      \csname LTb\endcsname%
      \put(2534,1022){\makebox(0,0)[r]{\strut{}}}%
      \csname LTb\endcsname%
      \put(2534,1339){\makebox(0,0)[r]{\strut{}}}%
      \csname LTb\endcsname%
      \put(2534,1656){\makebox(0,0)[r]{\strut{}}}%
      \csname LTb\endcsname%
      \put(2534,1973){\makebox(0,0)[r]{\strut{}}}%
      \csname LTb\endcsname%
      \put(2903,2244){\makebox(0,0){\strut{} 1}}%
      \csname LTb\endcsname%
      \put(3379,2244){\makebox(0,0){\strut{} 2}}%
      \csname LTb\endcsname%
      \put(3855,2244){\makebox(0,0){\strut{} 3}}%
      \csname LTb\endcsname%
      \put(4331,2244){\makebox(0,0){\strut{} 4}}%
    }%
    \gplbacktext
\put(0,0){\includegraphics[width=8.4cm]{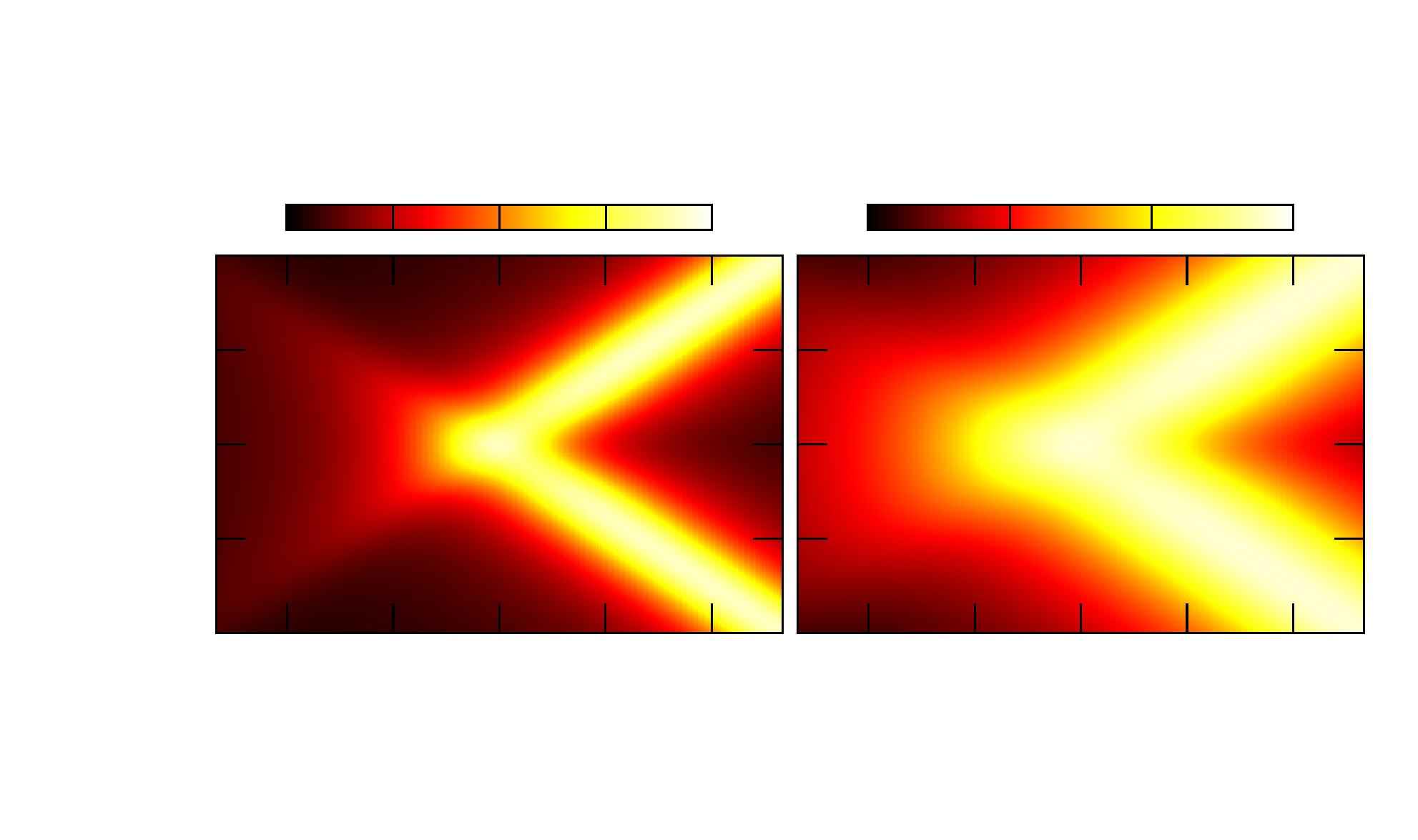}}
    \gplfronttext
  \end{picture}%
\endgroup
\caption{\label{fig:noise:wbs} (Color online) Case of a good conductor: Density plots of the ratio 
$\mathcal{S}_R(\omega,\omega_{\mathrm{dc}})$ of the backscattering
current noise for $R\neq 0$ to the same quantity at $R=0$ (no environment)
as function of $\hbar\omega/E_B$ and $\hbar\omega_{\mathrm{dc}}/E_B$ for $R=2R_q$ ($K=1/3$) and $R=R_q/2$ ($K=2/3$)
assuming $k_BT=5\,E_B$. Frequencies $\omega$ and
	$\omega_{\mathrm{dc}}$ are
kept below the high energy cutoff $\hbar\omega_{\mathrm{RC}}=40\,E_B$.
}
	\end{center} 
\end{figure}

Figure~\ref{fig:noise:wbs} presents $\mathcal{S}_R(\omega,\omega_{\mathrm{dc}})$ for a good
conductor, assuming $k_BT/E_B=5$.
In this regime, $\mathcal{S}_R(\omega,\omega_{\mathrm{dc}})\geq 1$: the environment
enhances the backscattering noise. This enhancement is especially strong in the thermal regime
and for the emission noise within the diagonal bands $|\omega-|\omega_{\text{dc}}||\lesssim k_BT/\hbar$.
It becomes even stronger and more concentrated along the 
diagonal bands with increasing $R$. 
This apparently surprising
result comes from the fact that increasing $R$ leads to a stronger DCB of
the total current
corresponding to an increase of $I_B$ and correspondingly of its FF noise.

\begin{figure}[htbp] 
	\begin{center}
\begingroup
  \makeatletter
  \providecommand\color[2][]{%
    \GenericError{(gnuplot) \space\space\space\@spaces}{%
      Package color not loaded in conjunction with
      terminal option `colourtext'%
    }{See the gnuplot documentation for explanation.%
    }{Either use 'blacktext' in gnuplot or load the package
      color.sty in LaTeX.}%
    \renewcommand\color[2][]{}%
  }%
  \providecommand\includegraphics[2][]{%
    \GenericError{(gnuplot) \space\space\space\@spaces}{%
      Package graphicx or graphics not loaded%
    }{See the gnuplot documentation for explanation.%
    }{The gnuplot epslatex terminal needs graphicx.sty or graphics.sty.}%
    \renewcommand\includegraphics[2][]{}%
  }%
  \providecommand\rotatebox[2]{#2}%
  \@ifundefined{ifGPcolor}{%
    \newif\ifGPcolor
    \GPcolortrue
  }{}%
  \@ifundefined{ifGPblacktext}{%
    \newif\ifGPblacktext
    \GPblacktexttrue
  }{}%
  \let\gplgaddtomacro\g@addto@macro
  \gdef\gplbacktext{}%
  \gdef\gplfronttext{}%
  \makeatother
  \ifGPblacktext
    \def\colorrgb#1{}%
    \def\colorgray#1{}%
  \else
    \ifGPcolor
      \def\colorrgb#1{\color[rgb]{#1}}%
      \def\colorgray#1{\color[gray]{#1}}%
      \expandafter\def\csname LTw\endcsname{\color{white}}%
      \expandafter\def\csname LTb\endcsname{\color{black}}%
      \expandafter\def\csname LTa\endcsname{\color{black}}%
      \expandafter\def\csname LT0\endcsname{\color[rgb]{1,0,0}}%
      \expandafter\def\csname LT1\endcsname{\color[rgb]{0,1,0}}%
      \expandafter\def\csname LT2\endcsname{\color[rgb]{0,0,1}}%
      \expandafter\def\csname LT3\endcsname{\color[rgb]{1,0,1}}%
      \expandafter\def\csname LT4\endcsname{\color[rgb]{0,1,1}}%
      \expandafter\def\csname LT5\endcsname{\color[rgb]{1,1,0}}%
      \expandafter\def\csname LT6\endcsname{\color[rgb]{0,0,0}}%
      \expandafter\def\csname LT7\endcsname{\color[rgb]{1,0.3,0}}%
      \expandafter\def\csname LT8\endcsname{\color[rgb]{0.5,0.5,0.5}}%
    \else
      \def\colorrgb#1{\color{black}}%
      \def\colorgray#1{\color[gray]{#1}}%
      \expandafter\def\csname LTw\endcsname{\color{white}}%
      \expandafter\def\csname LTb\endcsname{\color{black}}%
      \expandafter\def\csname LTa\endcsname{\color{black}}%
      \expandafter\def\csname LT0\endcsname{\color{black}}%
      \expandafter\def\csname LT1\endcsname{\color{black}}%
      \expandafter\def\csname LT2\endcsname{\color{black}}%
      \expandafter\def\csname LT3\endcsname{\color{black}}%
      \expandafter\def\csname LT4\endcsname{\color{black}}%
      \expandafter\def\csname LT5\endcsname{\color{black}}%
      \expandafter\def\csname LT6\endcsname{\color{black}}%
      \expandafter\def\csname LT7\endcsname{\color{black}}%
      \expandafter\def\csname LT8\endcsname{\color{black}}%
    \fi
  \fi
  \setlength{\unitlength}{0.0500bp}%
  \begin{picture}(4760.00,2820.00)%
    \gplgaddtomacro\gplbacktext{%
      \csname LTb\endcsname%
      \put(1714,2587){\makebox(0,0){\strut{}$R = 2R_q$}}%
    }%
    \gplgaddtomacro\gplfronttext{%
      \csname LTb\endcsname%
      \put(953,463){\makebox(0,0){\strut{}-0.8}}%
      \csname LTb\endcsname%
      \put(1334,463){\makebox(0,0){\strut{}-0.4}}%
      \csname LTb\endcsname%
      \put(1714,463){\makebox(0,0){\strut{} 0}}%
      \csname LTb\endcsname%
      \put(2094,463){\makebox(0,0){\strut{} 0.4}}%
      \csname LTb\endcsname%
      \put(2475,463){\makebox(0,0){\strut{} 0.8}}%
      \csname LTb\endcsname%
      \put(1714,184){\makebox(0,0){\strut{}$\hbar \omega / E_B$}}%
      \csname LTb\endcsname%
      \put(630,705){\makebox(0,0)[r]{\strut{}-1}}%
      \csname LTb\endcsname%
      \put(630,1022){\makebox(0,0)[r]{\strut{}-0.5}}%
      \csname LTb\endcsname%
      \put(630,1339){\makebox(0,0)[r]{\strut{} 0}}%
      \csname LTb\endcsname%
      \put(630,1656){\makebox(0,0)[r]{\strut{} 0.5}}%
      \csname LTb\endcsname%
      \put(630,1973){\makebox(0,0)[r]{\strut{} 1}}%
      \csname LTb\endcsname%
      \put(141,1339){\rotatebox{-270}{\makebox(0,0){\strut{}$\hbar \omega_{\mathrm{dc}} / E_B$}}}%
      \csname LTb\endcsname%
      \put(999,2244){\makebox(0,0){\strut{} 0}}%
      \csname LTb\endcsname%
      \put(1599,2244){\makebox(0,0){\strut{}2e-9}}%
      \csname LTb\endcsname%
      \put(2199,2244){\makebox(0,0){\strut{}4e-9}}%
    }%
    \gplgaddtomacro\gplbacktext{%
      \csname LTb\endcsname%
      \put(3665,2587){\makebox(0,0){\strut{}$R = R_q/2$}}%
    }%
    \gplgaddtomacro\gplfronttext{%
      \csname LTb\endcsname%
      \put(2904,463){\makebox(0,0){\strut{}-0.8}}%
      \csname LTb\endcsname%
      \put(3285,463){\makebox(0,0){\strut{}-0.4}}%
      \csname LTb\endcsname%
      \put(3665,463){\makebox(0,0){\strut{} 0}}%
      \csname LTb\endcsname%
      \put(4045,463){\makebox(0,0){\strut{} 0.4}}%
      \csname LTb\endcsname%
      \put(4426,463){\makebox(0,0){\strut{} 0.8}}%
      \csname LTb\endcsname%
      \put(3665,184){\makebox(0,0){\strut{}$\hbar \omega / E_B$}}%
      \csname LTb\endcsname%
      \put(2581,705){\makebox(0,0)[r]{\strut{}}}%
      \csname LTb\endcsname%
      \put(2581,1022){\makebox(0,0)[r]{\strut{}}}%
      \csname LTb\endcsname%
      \put(2581,1339){\makebox(0,0)[r]{\strut{}}}%
      \csname LTb\endcsname%
      \put(2581,1656){\makebox(0,0)[r]{\strut{}}}%
      \csname LTb\endcsname%
      \put(2581,1973){\makebox(0,0)[r]{\strut{}}}%
      \csname LTb\endcsname%
      \put(2951,2244){\makebox(0,0){\strut{} 0}}%
      \csname LTb\endcsname%
      \put(3665,2244){\makebox(0,0){\strut{} 0.008}}%
      \csname LTb\endcsname%
      \put(4379,2244){\makebox(0,0){\strut{} 0.016}}%
    }%
    \gplbacktext
	\put(0,0){\includegraphics[width=8.4cm]{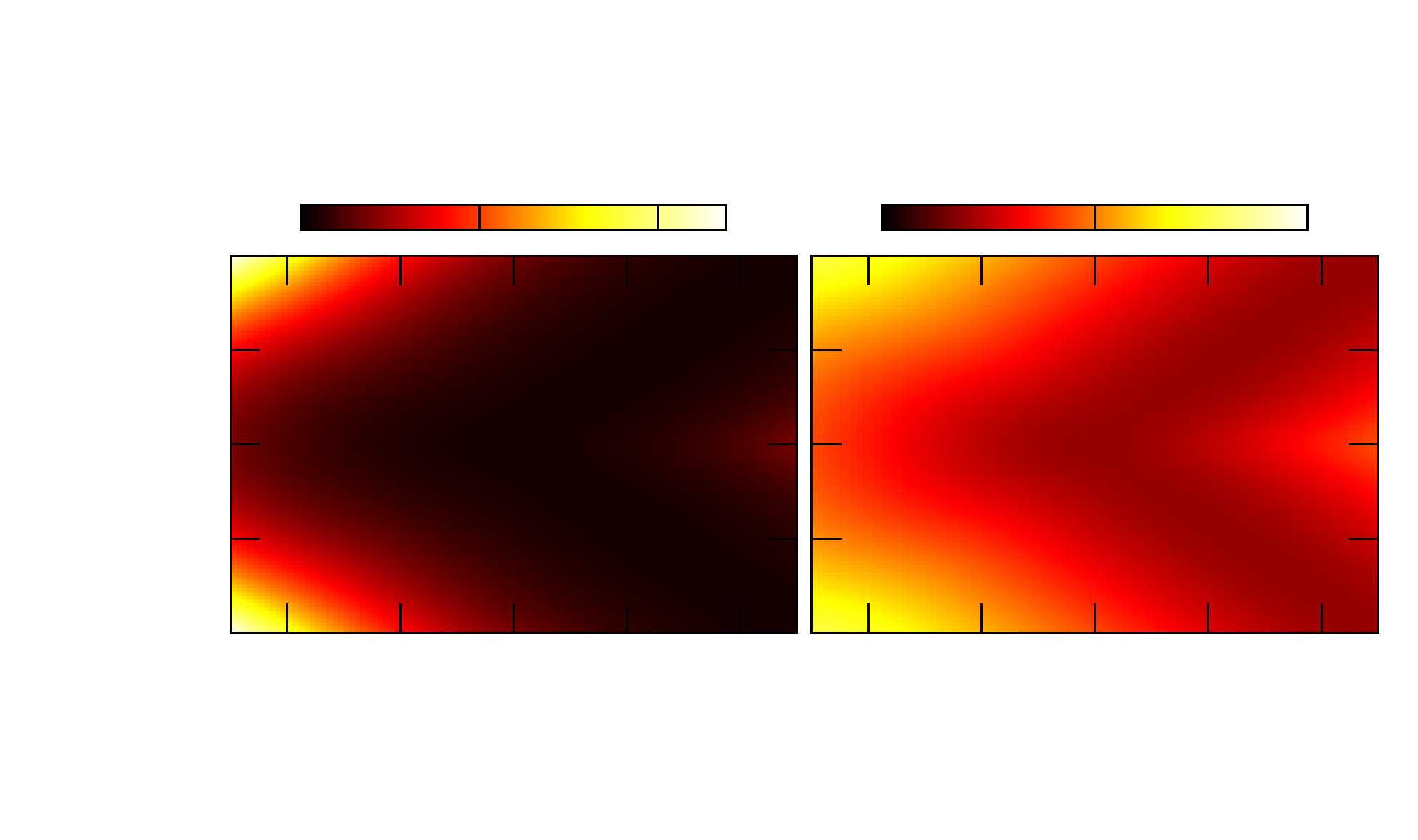}}
    \gplfronttext
  \end{picture}%
\endgroup
\caption{ \label{fig:noise:sbs} (Color online) Weakly transmitting
regime: Density plot of the ratio 
$\mathcal{S}_R(\omega,\omega_{\mathrm{dc}})$ of the transmitted 
current noise for $R\neq 0$ to the same quantity at $R=0$ (no environment)
as function of $\hbar\omega/E_B$ and $\hbar\omega_{\mathrm{dc}}/E_B$ for 
$R=2R_q$ ($K=1/3$) and $R=R_q/2$ ($K=2/3$)
for $\hbar\omega_{\text{RC}}=40\,E_B$ and $k_BT=E_B/10$. The plot is limited to energy scales below
$E_B$ to explore the vicinity of the IR fixed point of the TLL model. 
		}
	\end{center} 
\end{figure}

Figure~\ref{fig:noise:sbs} presents $\mathcal{S}_R(\omega,\omega_{\mathrm{dc}})$ in the strong backscattering 
regime describing both a weakly transmitting conductor and the low energy behavior of 
a good conductor. At low temperature
$k_BT=E_B/10$, a DCB of the noise 
is observed as expected from the DCB of the DC non-equilibrium current.

Based on a generalized mapping between the DCB problem and a generalized 
TLL model~\cite{Jezouin:2013-1,Souquet:2013-1}, we expect
these conclusions on noise enhancement/reduction by environmental
effect to remain qualitatively valid for a linear environment with 
a frequency dependent impedance. 

\medskip

To conclude, we have obtained a non-equilibrium perturbative FDR relating
the FF noise to the non-equilibrium DC current across a generic two-terminal
quantum circuit. This model independent FDR unifies many previous results
and only relies on three
hypotheses among which a detailed balance condition in the limit of vanishing tunneling.
Most importantly, this condition is deeply related to
the effective thermalization of the whole circuit,
a question of first importance in mesoscopic thermodynamics~\cite{Sanchez:2010-1}.
This out of equilibrium FDR opens the way to numerous
experiments: first of all, testing it on complex
nano-structures such as quantum dots would check the basic hypotheses
\textit{(i)—(iii)}. 
Our FDR can also be used to determine whether one measures the symmetrized
or non-symmetrized noise when accessing only the emission
part of the noise spectrum ($\omega>0$)~\cite{Altimiras:2014-1}.
Finally, our non-equilibrium FDR and its generalization to AC
bias~\cite{Safi:2014-1} provide complementary methods to
measure the effective tunneling charge $q$ in addition to these proposed in
Ref.~\cite{Safi:2010-1}.

\begin{acknowledgments}
I.S thanks E.~Sukhorukov for previous
collaboration and useful insight.
We also thank C. Altimiras, D.~Est\`eve, 
P.~Joyez, F.~Portier for useful remarks on this manuscript as well as
D.G.~Glattli, F.~Hekking and B.~Reulet for stimulating discussions.
This work was partially supported by 
the ANR grant ``1shot reloaded''(ANR-14-CE32-0017).
\end{acknowledgments}


\appendix
\section{Derivation of the Fluctuation Dissipation Relation}
\label{appendix:FDR}

The first step is to compute, to the lowest relevant order in perturbation theory, both 
the average current $\langle I(t)\rangle $ and 
current correlations defined in the time domain by
\begin{equation}
\label{eq:defnoise}
S_I(t,t') = \langle I(t)\,I(t')\rangle -\langle I(t)\rangle\langle
I(t')\rangle
\end{equation}
Eq.~\eqref{eq:defnoise} defines 
a quantum (non-symmetrized) correlation function which, in full generality, 
is not symmetric with respect to $t\leftrightarrow t'$. 
The non-symmetrized current noise at
finite frequency is then defined by
\begin{equation}
\label{eq:defnoise:ff}
S_I(\omega)= \int_{-\infty}^{+\infty} \me^{\mi\omega \tau}\overline{
S_I\left(\bar{t}-\frac{\tau}{2},\bar{t}+\frac{\tau}{2}\right)}^{\bar{t}}
\,\mathrm{d}\tau
\end{equation}
where $\overline{S_I(\bar{t}-\tau/2,\bar{t}+\tau/2)}^{\bar{t}}$ denotes the time average over
$\bar{t}=(t+t')/2$ of $S_I(t,t')$ arising from the long acquisition time of the
noise signal. With this definition,  
the emission noise corresponds to $\omega>0$ whereas the absorption 
noise corresponds to $\omega<0$.
Generic current
noise measurement corresponds to emission noise 
measurements~\cite{Gavish:2003-1} but 
it is now possible to access the full
non-symmetrized noise by exploiting the photo-assisted
tunneling of quasi-particles across an
on-chip superconductor-insulator-superconductor 
junction~\cite{Deblock:2003-1,Billangeon:2006-1,Basset:2010-1,Basset:2012-1}.
In the recently performed experiments on the dynamical Coulomb blockade of
the noise~\cite{Altimiras:2014-1,Parlavecchio:2015-1}, the detection setup accesses 
the non-symmetrized excess
noise due to the AC bias of the quantum circuit even if the
final detection stage is a standard power measurement. 

\subsection{Perturbation theory}

Results are most easily obtained by going into the interation representation with respect to the
Hamlitlonien $\mathcal{H}_0$ (no tunneling) and expand the evolution operators 
in powers of the operators $\mathcal{A}_0(t')$ and
$\mathcal{A}_0^\dagger(t')$ 
which include environmental phases, expressed in the
interaction scheme with respect to the Hamiltonian $\mathcal{H}_0$.

\medskip

\paragraph{The average current}
The zero-th order term trivially vanishes and,
at the lowest non-trivial order, half of the terms involve either two
operators $\mathcal{A}$ or two operators $\mathcal{A}^\dagger$. These terms
are assumed to vanish
due to hypothesis (ii). The remaining terms involve exactly one $\mathcal{A}$ 
and one $\mathcal{A}^\dagger$ operator. They can therefore be expressed in terms of the two correlators:
\begin{subequations}
\begin{eqnarray}
X^>(t,t') & = & \mathrm{Tr}(\mathcal{A}_0(t)\,\rho_0\,\mathcal{A}_0^\dagger(t'))\\
X^<(t,t') & = & \mathrm{Tr}(\mathcal{A}^\dagger_0(t)\,\rho_0\,\mathcal{A}_0(t')).
\end{eqnarray}
\end{subequations}
Consequently, 
at lowest order, under our hypotheses, the average current at time $t$ is obtained as 
\begin{subequations}
\label{eq:current:suppl}
\begin{align}
\langle I(t)\rangle & = \frac{q}{\hbar^2}\int_0^tE(t') E(t)^*(X^<(t',t)-X^>(t,t'))\,\mathrm{d}t'\\
 & + \frac{q}{\hbar^2} \int_0^t E(t')^* E(t) (X^>(t,t')-X^<(t,t'))\,\mathrm{d}t'\,.
\end{align}
\end{subequations}

\paragraph{Current noise}
Let us now turn to the current noise. In this case, the second order
in perturbation theory is obtained
without expanding the evolution operators since we have already products
of two operators $\mathcal{A}$ and $\mathcal{A}^\dagger$. Since the product $\langle
I(t)\rangle \langle I(t')\rangle$ is at fourth order,
the current noise is finally given by:
\begin{equation}
\label{eq:general}
S_I(t,t')=\frac{q^2}{\hbar}\left( E(t)^*E(t')\,X^>(t',t) +
E(t')^*E(t)\,X^<(t',t)\right)\,.
\end{equation}
Although these relations are valid for a general time dependent voltage and
are thus relevant for discussing photo-assisted noise and current, we
shall now focus on the stationary regime where $V(t)=V$ and therefore
$E(t)\propto e^{-i\omega_{\text{dc}}t}$.
As we shall see now, the above expressions will
simplify and provide an explicit FDR relating the noise spectrum of the
quantum noise to the non-equilibrium dc characteristic of the conductor.

\subsection{Stationary case}

In the stationary case, the tunneling correlators $X^>(t,t')$ and $X^<(t,t')$ only depend 
on the difference $\tau=t-t'$. Introducing the Fourier transforms
\begin{subequations}
\begin{eqnarray}
X^>(\omega) & = & \int_{-\infty}^{+\infty}
X^>\left(\frac{\tau}{2},-\frac{\tau}{2}\right)\,\me^{\mi\omega\tau}\md\tau
\\
X^<(\omega) & = & \int_{-\infty}^{+\infty}
X^<\left(\frac{\tau}{2},-\frac{\tau}{2}\right)\,\me^{\mi\omega\tau}\md\tau\,,
\end{eqnarray}
\end{subequations}
and using Eqs.~\eqref{eq:current:suppl},
the non-equilibrium DC current is obtained as ($\omega_{\text{dc}}=qV/\hbar$):
\begin{equation}
\label{eq:ff:current}
I(\omega_{\mathrm{dc}})= \frac{q}{\hbar^2}\left( 
X^<(-\omega_{\mathrm{dc}})-X^>(\omega_{\mathrm{dc}})\right)\,.
\end{equation}
Substituting Eq.~\eqref{eq:defnoise:ff} into Eq.~\eqref{eq:general} then leads to
the FF noise:
\begin{equation}
\label{eq:ff:noise}
S_I(\omega)= \frac{q^2}{\hbar^2}\left(
X^<(\omega-\omega_{\text{dc}})+X^>(\omega+\omega_{\text{dc}})\right)\,.
\end{equation}
The detailed balance (hypothesis (iii)) relates the occupation probabilities $p_I$
of the many body eigenstates $|I\rangle$ of the circuit with energies $E_I$
in the limit of vanishing tunneling:
\begin{equation}
\label{eq:db}
\frac{p_I}{p_J}=\me^{-(E_I-E_J)/k_BT}\,.
\end{equation}
Using \eqref{eq:db} within the K\"{a}llen-Lehmann spectral representation of
$X^>(t,t')$ and
$X^<(t,t')$, the Fourier transforms $X^>(\omega)$ and $X^<(-\omega)$ can be related through
\begin{equation}
X^<(-\omega)=\me^{\hbar\omega/k_BT}X^>(\omega)\,.
\end{equation}
This enables us to express $X^>(\omega)$ as well as $X^<(-\omega)$ 
in terms of the dc out of equilibrium current:
\begin{subequations}
\label{eq:Keldysh}
\begin{eqnarray}
q\,X^>(\omega) & = & \hbar^2 N(\omega)\,I(\omega)\\
q\,X^<(-\omega) & = & \hbar^2(N(\omega)+1)\,I(\omega)
\end{eqnarray}
\end{subequations}
where $N(\omega)=(\me^{\hbar\omega/k_BT}-1)^{-1}$ denotes the Bose occupation
number and, in the above expressions, this expression is also used to
define $N(\omega)$ 
$\omega<0$. Substituting 
Eq.~\eqref{eq:Keldysh} into Eq.~\eqref{eq:ff:noise} leads to our main
result, \textit{i.e.} the expression of the FF noise in terms of the out of
equilibrium dc current:
\begin{align}
\label{eq:result}
S_I(\omega) & = q\left(N(\omega_{\text{dc}}+\omega) I(\omega_{\text{dc}}+\omega) \right. \nonumber\\
& \left. +(1+N(\omega_{\text{dc}}-\omega)) I(\omega_{\text{dc}}-\omega)\right)\,.
\end{align}

\section{The Tomonaga Luttinger barrier problem}
\label{appendix:TLL}

\subsection{Presentation of the problem}

The problem of a single localized barrier in the TLL has been originally
studied by Kane and Fisher within a renormalization group
approach~\cite{Kane:1992-1}. In the interacting case ($K\neq 1$), their work has revealed a phase diagram
showing an UV and an IR fixed point respectively corresponding to a fully transmitting
conductor and a disconnected conductor. These fixed points exchange their
stability between the repulsive case $K<1$ and the attractive case $K>1$.
In the repulsive case, the UV fixed point is unstable whereas the IR fixed
point is stable, thus implying that the effective barrier diverges at low
energies whereas it vanishes at low energies.

\medskip

A full solution of the problem has been provided in the non-equilibrium
stationary case using the thermodynamical Bethe
ansatz technique, thus allowing a full interpolation between these two
fixed points and explicit predictions for the non-equilibrium current and
the low frequency noise~\cite{Fendley:1995-1,Fendley:1995-2} with natural
applications to the fractional quantum Hall effect (FQHE)~\cite{Fendley:1995-3}.
Further work has led to the determination of the full counting statistics
of the charge flowing across a quantum point contact in the
FQHE~\cite{Saleur:2001-1} and in the TLL~\cite{Komnik:2006-1}. 
More recently, an exact description of non-equilibrium fixed points of
quantum impurity models suitable for treating time-dependant problems has
been proposed~\cite{Freton:2014-1} and may open a non-perturbative approach
to generalize or go beyond the perturbative results discussed here.

\medskip

In the present letter, we shall only use perturbative results for
the out of equilibrium current in the vicinity of
both fixed points. Such explicit perturbative expressions have been obtained for
the non-equilibrium current in the vicinity of the UV fixed point by Chamon
{\it et al}~\cite{Chamon:1995-1} but the duality of the local barrier
problem in a TLL~\cite{Fendley:1998-1,Fendley:1998-2} 
enables us to use similar expressions close to
the IR fixed point provided one replaces $K$ by $1/K$. 

\subsection{The weak backscattering regime}

In this regime, the barrier is modeled by a localized potential described
by a localized potential $v_B$. The model also has a high frequency cutoff
denoted here by $\omega_{\text{RC}}$.
A relation between microscopic parameters $v_B$ and $\omega_{\text{RC}}$
and measurable quantities is
obtained by considering $K=1$ which corresponds to the bare conductor
($R=0$ in the DCB problem). Then denoting $\mathcal{T}=G_{K=1}(T)$ which
indeed does not depend anymore on the temperature, we find:
\begin{equation}
\label{eq:TvBrelation}
\mathcal{T}=1-\frac{(\pi v_B)^2}{\omega_{\text{RC}}}
\end{equation}
Let us now consider $K<1$ directly relevant for the weak
backscattering regime since $K=(1+R/R_q)^{-1}$~\cite{Safi:2004-1}.

\medskip

At first non-trivial order in perturbation theory, the total current flowing across the barrier 
contains a weak backscattering correction to the bare current $q^2K V/h$:  
\begin{equation}
I(V,T) =
\frac{e^2}{h}V\left(K-G_K(T)\,\mathcal{F}_K\left(\frac{qV}{k_BT}\right)\right)\,
\end{equation}
where 
\begin{equation}
\label{eq:Fdefinition}
\mathcal{F}_K(z)
=\frac{\sinh{(z/2)}}{z/2}\,\frac{\left|\Gamma\left(K+
\frac{\mi z}{2\pi} \right)\right|^2}{\Gamma(K)^2}
\end{equation}
and $G_K(T)$ is a dimensionless backscattering conductance whose
expression depends on the cutoff as well as on the potential of the
barrier:
\begin{equation}
G_K(T) = \pi^2
\frac{\Gamma(K)^2}{\Gamma(2K)}\,\frac{v_B^2}{\omega_{\text{RC}}}\left(
\frac{\hbar\omega_{\text{RC}}}{\pi k_BT}\right)^{2(1-K)}
\end{equation}
Using Eq.~\eqref{eq:TvBrelation}, we can then reexpress the dimensionless backscattering linear conductance
$G_K(T)$ in 
presence of interactions in terms of the temperature and of the energy
scale $E_B$ associated with the barrier:
\begin{equation}
\label{eq:GK}
G_K(T)= \frac{\Gamma(K)^2}{\Gamma(2K)}\,\left(
\frac{E_B}{\pi_BT}\right)^{2(1-K)}
\end{equation}
where $E_B$ is related to the microscopic parameters
through the relation
\begin{equation}
E_B=\hbar\omega_{\text{RC}}\,(1-\mathcal{T})^{1/2(1-K)}\,.
\end{equation} 
valid in the limit $\mathcal{T}\sim 1$.

\subsection{The strong backscattering regime}

In the strong backscattering regime, the system is modeled by two half
infinite 1D TLLs coupled by a tunneling barrier described by a tunneling amplitude $\Gamma$.
At the IR fixed point ($\Gamma=0$), no current flows across the barrier.
When switching on tunneling, current can flow but, at zero temperature, interactions lead to a
non-linear non-equilibrium current in terms of the dc
bias~\cite{Kane:1992-1}. At non zero temperature, a temperature dependent
linear conductance can still be defined and vanishes with temperature/
Nevertheless, second order perturbation theory in $\Gamma$
leads to the following expression for the non-equilibrium
current:
\begin{equation}
I(V,T) = 
\frac{e^2}{h}VG_{1/K}\,(T)\mathcal{F}_{1/K}\left(\frac{qV}{k_BT}\right)\,
\end{equation}
where $\mathcal{F}_{1/K}(z)$ is given by Eq.~\eqref{eq:Fdefinition} remplacing $K$ by
$1/K$ and 
$G_{1/K}(T)$ denotes the linear conductance at finite temperature $T$ whose
expression is given by substituting $K$ by $1/K$ and $v_B$ by $\Gamma$ in Eq.~\eqref{eq:GK}. As
expected, $G_{1/K}(T)$ vanishes at low temperature for $K<1$.

\subsection{Limitations to the perturbative approach for a good conductor}

Here we discuss how the breakdown of perturbation theory manifests itself
in predictions of the FF noise deduced from Eq.~\eqref{eq:result}.
Let us consider here the
case of a good conductor, keeping $|\hbar\omega|$ and $|qV|$ smaller than
the high energy cutoff
$\hbar\omega_{\text{RC}}$.

\medskip

\begin{figure*}[htbp] 
\centering
\begingroup
  \makeatletter
  \providecommand\color[2][]{%
    \GenericError{(gnuplot) \space\space\space\@spaces}{%
      Package color not loaded in conjunction with
      terminal option `colourtext'%
    }{See the gnuplot documentation for explanation.%
    }{Either use 'blacktext' in gnuplot or load the package
      color.sty in LaTeX.}%
    \renewcommand\color[2][]{}%
  }%
  \providecommand\includegraphics[2][]{%
    \GenericError{(gnuplot) \space\space\space\@spaces}{%
      Package graphicx or graphics not loaded%
    }{See the gnuplot documentation for explanation.%
    }{The gnuplot epslatex terminal needs graphicx.sty or graphics.sty.}%
    \renewcommand\includegraphics[2][]{}%
  }%
  \providecommand\rotatebox[2]{#2}%
  \@ifundefined{ifGPcolor}{%
    \newif\ifGPcolor
    \GPcolortrue
  }{}%
  \@ifundefined{ifGPblacktext}{%
    \newif\ifGPblacktext
    \GPblacktexttrue
  }{}%
  \let\gplgaddtomacro\g@addto@macro
  \gdef\gplbacktext{}%
  \gdef\gplfronttext{}%
  \makeatother
  \ifGPblacktext
    \def\colorrgb#1{}%
    \def\colorgray#1{}%
  \else
    \ifGPcolor
      \def\colorrgb#1{\color[rgb]{#1}}%
      \def\colorgray#1{\color[gray]{#1}}%
      \expandafter\def\csname LTw\endcsname{\color{white}}%
      \expandafter\def\csname LTb\endcsname{\color{black}}%
      \expandafter\def\csname LTa\endcsname{\color{black}}%
      \expandafter\def\csname LT0\endcsname{\color[rgb]{1,0,0}}%
      \expandafter\def\csname LT1\endcsname{\color[rgb]{0,1,0}}%
      \expandafter\def\csname LT2\endcsname{\color[rgb]{0,0,1}}%
      \expandafter\def\csname LT3\endcsname{\color[rgb]{1,0,1}}%
      \expandafter\def\csname LT4\endcsname{\color[rgb]{0,1,1}}%
      \expandafter\def\csname LT5\endcsname{\color[rgb]{1,1,0}}%
      \expandafter\def\csname LT6\endcsname{\color[rgb]{0,0,0}}%
      \expandafter\def\csname LT7\endcsname{\color[rgb]{1,0.3,0}}%
      \expandafter\def\csname LT8\endcsname{\color[rgb]{0.5,0.5,0.5}}%
    \else
      \def\colorrgb#1{\color{black}}%
      \def\colorgray#1{\color[gray]{#1}}%
      \expandafter\def\csname LTw\endcsname{\color{white}}%
      \expandafter\def\csname LTb\endcsname{\color{black}}%
      \expandafter\def\csname LTa\endcsname{\color{black}}%
      \expandafter\def\csname LT0\endcsname{\color{black}}%
      \expandafter\def\csname LT1\endcsname{\color{black}}%
      \expandafter\def\csname LT2\endcsname{\color{black}}%
      \expandafter\def\csname LT3\endcsname{\color{black}}%
      \expandafter\def\csname LT4\endcsname{\color{black}}%
      \expandafter\def\csname LT5\endcsname{\color{black}}%
      \expandafter\def\csname LT6\endcsname{\color{black}}%
      \expandafter\def\csname LT7\endcsname{\color{black}}%
      \expandafter\def\csname LT8\endcsname{\color{black}}%
    \fi
  \fi
  \setlength{\unitlength}{0.0500bp}%
  \begin{picture}(9960.00,2820.00)%
    \gplgaddtomacro\gplbacktext{%
      \csname LTb\endcsname%
      \put(1593,2535){\makebox(0,0){\strut{}(a) $k_B T = 2\, E_B$}}%
    }%
    \gplgaddtomacro\gplfronttext{%
      \csname LTb\endcsname%
      \put(697,605){\makebox(0,0){\strut{}-40}}%
      \csname LTb\endcsname%
      \put(1145,605){\makebox(0,0){\strut{}-20}}%
      \csname LTb\endcsname%
      \put(1593,605){\makebox(0,0){\strut{} 0}}%
      \csname LTb\endcsname%
      \put(2041,605){\makebox(0,0){\strut{} 20}}%
      \csname LTb\endcsname%
      \put(2489,605){\makebox(0,0){\strut{} 40}}%
      \csname LTb\endcsname%
      \put(1593,326){\makebox(0,0){\strut{}$\hbar \omega / E_B$}}%
      \csname LTb\endcsname%
      \put(565,847){\makebox(0,0)[r]{\strut{}-40}}%
      \csname LTb\endcsname%
      \put(565,1199){\makebox(0,0)[r]{\strut{}-20}}%
      \csname LTb\endcsname%
      \put(565,1551){\makebox(0,0)[r]{\strut{} 0}}%
      \csname LTb\endcsname%
      \put(565,1903){\makebox(0,0)[r]{\strut{} 20}}%
      \csname LTb\endcsname%
      \put(565,2255){\makebox(0,0)[r]{\strut{} 40}}%
      \csname LTb\endcsname%
      \put(158,1551){\rotatebox{-270}{\makebox(0,0){\strut{}$\hbar \omega_{\mathrm{dc}} / E_B$}}}%
      \csname LTb\endcsname%
      \put(2725,847){\makebox(0,0)[l]{\strut{} 0}}%
      \csname LTb\endcsname%
      \put(2725,1199){\makebox(0,0)[l]{\strut{} 20}}%
      \csname LTb\endcsname%
      \put(2725,1551){\makebox(0,0)[l]{\strut{} 40}}%
      \csname LTb\endcsname%
      \put(2725,1903){\makebox(0,0)[l]{\strut{} 60}}%
      \csname LTb\endcsname%
      \put(2725,2255){\makebox(0,0)[l]{\strut{} 80}}%
    }%
    \gplgaddtomacro\gplbacktext{%
      \csname LTb\endcsname%
      \put(4880,2535){\makebox(0,0){\strut{}(b) $k_B T = 0.5\, E_B$}}%
    }%
    \gplgaddtomacro\gplfronttext{%
      \csname LTb\endcsname%
      \put(3984,605){\makebox(0,0){\strut{}-40}}%
      \csname LTb\endcsname%
      \put(4432,605){\makebox(0,0){\strut{}-20}}%
      \csname LTb\endcsname%
      \put(4880,605){\makebox(0,0){\strut{} 0}}%
      \csname LTb\endcsname%
      \put(5328,605){\makebox(0,0){\strut{} 20}}%
      \csname LTb\endcsname%
      \put(5776,605){\makebox(0,0){\strut{} 40}}%
      \csname LTb\endcsname%
      \put(4880,326){\makebox(0,0){\strut{}$\hbar \omega / E_B$}}%
      \csname LTb\endcsname%
      \put(3852,847){\makebox(0,0)[r]{\strut{}-40}}%
      \csname LTb\endcsname%
      \put(3852,1199){\makebox(0,0)[r]{\strut{}-20}}%
      \csname LTb\endcsname%
      \put(3852,1551){\makebox(0,0)[r]{\strut{} 0}}%
      \csname LTb\endcsname%
      \put(3852,1903){\makebox(0,0)[r]{\strut{} 20}}%
      \csname LTb\endcsname%
      \put(3852,2255){\makebox(0,0)[r]{\strut{} 40}}%
      \csname LTb\endcsname%
      \put(3445,1551){\rotatebox{-270}{\makebox(0,0){\strut{}$\hbar \omega_{\mathrm{dc}} / E_B$}}}%
      \csname LTb\endcsname%
      \put(6012,847){\makebox(0,0)[l]{\strut{} 0}}%
      \csname LTb\endcsname%
      \put(6012,1199){\makebox(0,0)[l]{\strut{} 100}}%
      \csname LTb\endcsname%
      \put(6012,1551){\makebox(0,0)[l]{\strut{} 200}}%
      \csname LTb\endcsname%
      \put(6012,1903){\makebox(0,0)[l]{\strut{} 300}}%
      \csname LTb\endcsname%
      \put(6012,2255){\makebox(0,0)[l]{\strut{} 400}}%
    }%
    \gplgaddtomacro\gplbacktext{%
      \csname LTb\endcsname%
      \put(8167,2535){\makebox(0,0){\strut{}(c) $k_B T = 0.1\, E_B$}}%
    }%
    \gplgaddtomacro\gplfronttext{%
      \csname LTb\endcsname%
      \put(7271,605){\makebox(0,0){\strut{}-40}}%
      \csname LTb\endcsname%
      \put(7719,605){\makebox(0,0){\strut{}-20}}%
      \csname LTb\endcsname%
      \put(8167,605){\makebox(0,0){\strut{} 0}}%
      \csname LTb\endcsname%
      \put(8615,605){\makebox(0,0){\strut{} 20}}%
      \csname LTb\endcsname%
      \put(9063,605){\makebox(0,0){\strut{} 40}}%
      \csname LTb\endcsname%
      \put(8167,326){\makebox(0,0){\strut{}$\hbar \omega / E_B$}}%
      \csname LTb\endcsname%
      \put(7139,847){\makebox(0,0)[r]{\strut{}-40}}%
      \csname LTb\endcsname%
      \put(7139,1199){\makebox(0,0)[r]{\strut{}-20}}%
      \csname LTb\endcsname%
      \put(7139,1551){\makebox(0,0)[r]{\strut{} 0}}%
      \csname LTb\endcsname%
      \put(7139,1903){\makebox(0,0)[r]{\strut{} 20}}%
      \csname LTb\endcsname%
      \put(7139,2255){\makebox(0,0)[r]{\strut{} 40}}%
      \csname LTb\endcsname%
      \put(6732,1551){\rotatebox{-270}{\makebox(0,0){\strut{}$\hbar \omega_{\mathrm{dc}} / E_B$}}}%
      \csname LTb\endcsname%
      \put(9299,847){\makebox(0,0)[l]{\strut{} 0}}%
      \csname LTb\endcsname%
      \put(9299,1199){\makebox(0,0)[l]{\strut{} 1000}}%
      \csname LTb\endcsname%
      \put(9299,1551){\makebox(0,0)[l]{\strut{} 2000}}%
      \csname LTb\endcsname%
      \put(9299,1903){\makebox(0,0)[l]{\strut{} 3000}}%
      \csname LTb\endcsname%
      \put(9299,2255){\makebox(0,0)[l]{\strut{} 4000}}%
    }%
    \gplbacktext
	\put(0,0){\includegraphics[width=17.6cm]{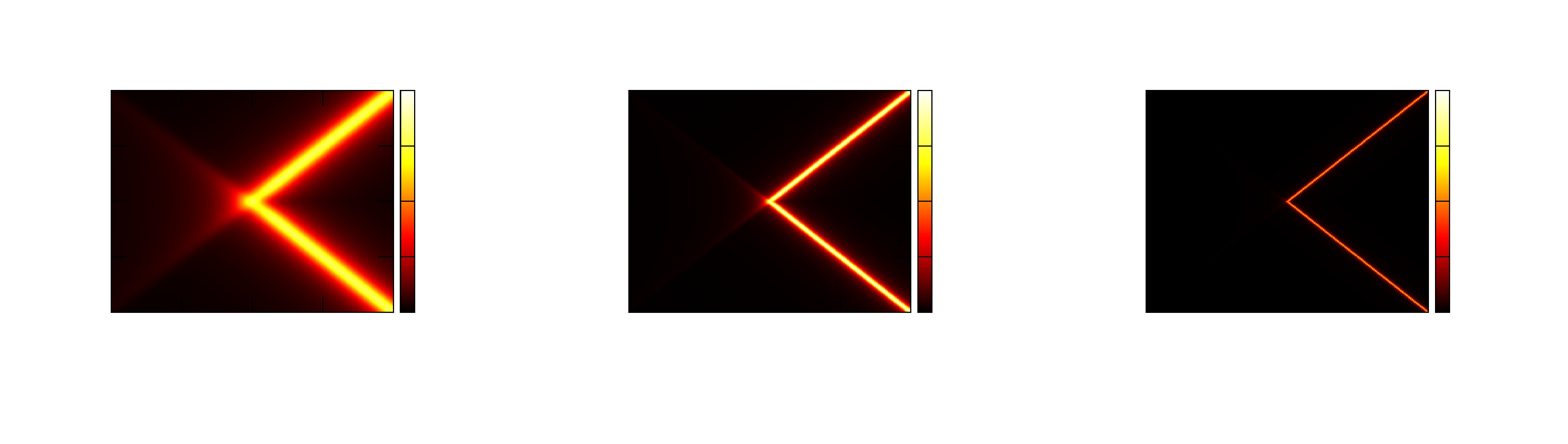}}%
    \gplfronttext
  \end{picture}%
\endgroup
\caption{\label{fig:noise:wbs:low-T} (Color online) Infrared divergences
for
a good conductor at low temperature: Density plots of the ratio 
$\mathcal{S}_R(\omega,\omega_{\mathrm{dc}})$ of the
backscattering
current noise for $R\neq 0$ to the same quantity at
$R=0$ (no environment)
as function of $\hbar\omega/E_B$ and
$\hbar\omega_{\mathrm{dc}}/E_B$ for $R=2R_q$
($K=1/3$) and various values of the temperature: (a) High temperature:
$k_BT=2\,E_B$, (b) Lower temperature: $k_BT=0.5\,E_B$ and (c) Lowest temperatures:
$k_BT=0.1\,E_B$. 
Frequencies $\omega$ and
$\omega_{\mathrm{dc}}$ are
kept below the high energy cutoff
$\hbar\omega_{\mathrm{RC}}=40\,E_B$. Note the different color scales of
each graph.
}
\end{figure*}

In order to see the breakdown of perturbation theory when one leaves the
vicinity of the UV fixed point of the TLL barrier problem, 
let us lower the
temperature, starting from $k_BT \gtrsim E_B$.
Fig.~\ref{fig:noise:wbs:low-T} depicts the prediction for the
dimensionless ratio 
\begin{equation}
\mathcal{S}_R(\omega,\omega_{\text{dc}})=\frac{S_{R,T}(\omega,\omega_{\text{dc}})}
{S_{R=0,T}(\omega,\omega_{\text{dc}})}
\end{equation}
associated
with the backscattering current for three
values of the temperature: $k_BT/E_B=2$, $0.5$ and $0.1$.

Fig.~\ref{fig:noise:wbs:low-T} clearly shows the signs of a divergence along in diagonal bands $|\omega -
|\omega_{\text{dc}}|| \lesssim E_B/\hbar$ when decreasing the temperature
below $E_B/k_B$. This is a signature
of the breakdown of perturbation theory since, in this region, when $k_BT$
becomes smaller than $E_B$, all energy scales involved in the
non-equilibrium current become smaller than $E_B$. This is precisely the
regime where perturbation theory close to the UV fixed point is expected to break down 
in the TLL barrier problem. Infrared divergences appearing along the diagonal
band expresses the divergence of the low energy fluctuations associated
with neutral excitations (e/h pairs).

\end{document}